\documentclass[aps,superscriptaddress,amsfonts,twocolumn,nofootinbib]{revtex4}



\usepackage{graphicx}
\usepackage{dcolumn}
\usepackage{tabularx}

\begin{document}

\title{Specialization of strategies and herding behavior of trading firms in a financial market}

\author{Fabrizio Lillo}
%\email[]{lillo@lagash.dft.unipa.it}
\affiliation{Dipartimento di Fisica e Tecnologie Relative, Universit\`a di Palermo,
Viale delle Scienze, I-90128, Palermo, Italy}
\affiliation{Santa Fe Institute, 1399 Hyde Park
Road, Santa Fe, NM 87501, USA}

\author{Esteban Moro}
%\email[]{}
\affiliation{Grupo Interdisciplinar de Sistemas Complejos (GISC), Departamento de Matem\'aticas, Universidad Carlos III de Madrid, Avenida de la Universidad 30, E-28911, Legan\'es, Spain}

\author{Gabriella Vaglica}
\affiliation{Dipartimento di Fisica e Tecnologie Relative, Universit\`a di Palermo,
Viale delle Scienze, I-90128, Palermo, Italy}
\affiliation{CNR-INFM, Unit\`a di Palermo, Palermo, Italy}

\author{Rosario N. Mantegna}
%\email[]{mantegna@unipa.it}
\affiliation{Dipartimento di Fisica e Tecnologie Relative, Universit\`a di Palermo,
Viale delle Scienze, I-90128, Palermo, Italy}

\date{\today}

\begin{abstract}
The understanding of complex social or economic systems is an important scientific challenge. Here we present a comprehensive study of the Spanish Stock Exchange showing that most financial firms trading in that market are characterized by a resulting strategy and can be classified in groups of firms with different specialization. Few large firms overally act as trending firms whereas many heterogeneous firm act as reversing firms. The herding properties of these two groups are markedly different and consistently observed over a four-year period of trading.
\end{abstract}

%\pacs{89.75.-k, 05.45.Tp, 02.10.Ox, 89.65.Gh}
%
\maketitle

{\it Introduction} --  The modeling of complex systems \cite{Simon1969,Anderson1972} benefits from the study of agent based models. A particular interesting system is the financial market. Despite many agent based models have been investigated \cite{Gode1993,Arthur1994,Levy1994,Lux1999,LeBaron1999,LeBaron2000,Tesfatsion2001,Challet2005}, only in few cases \cite{Lakonishok1992, Hardle1995,Nofsinger1999,Choe1999,Grinblatt2000,Griffin2003,Kossinets2006} an empirical investigation of agent strategies has been possible due to the lack of accessible data. The detection of empirical regularities in the investment strategy and herding behavior of large organizations may turn out to be a key advance in the design of empirically grounded agent based models of financial markets.

Complexity in financial markets emerges from the interaction at different levels of many individuals and organizations. There are institutional and individual investors taking investment decisions and their decisions are in most cases executed by financial and brokerage firms which are allowed to trade in a specific market.  Recent studies have empirically shown that the dynamics of institutional and individual trading can show detectable statistical regularities in investment decisions down to a daily or intradaily time horizon \cite{Griffin2003}. Broad classes of investment strategies such as the momentum \cite{Grinblatt1985} and the contrarian \cite{Chan1988} strategies have been empirically investigated, most frequently at quarterly intervals, by considering specialized databases allowing to track the investment decisions of large investors \cite{Lakonishok1992} and of both large and individual investors \cite{Nofsinger1999,Choe1999,Grinblatt2000,Griffin2003}. Momentum investors are buying stocks that were past winners. A contrarian strategy consists of buying stocks that have been loosers (or selling short stocks that have been winners). %The strategy is formulated on the assumption that the stock market overreacts and a contrarian investor can exploit the inefficiency related to market overreaction by reverting stock prices to fundamental values. 
Empirical investigations on different markets have shown that institutional investors are preferentially momentum investors whereas individual investors usually prefer a contrarian strategy.  Proprietary trading data obtained from the Korean Stock Exchange \cite{Choe1999} and from NASDAQ \cite{Griffin2003} have shown that stock returns have some ability to forecast inventory variation of groups of investors whereas the evidence of return predictability on the basis of investor inventory variation is negligible both at a daily and intradaily time horizon.

In the present study we empirically investigate the presence of detectable resulting investment strategies at the level of firms entitled to trade in a financial market. The investigated market is the Spanish Stock Market. 
In this market, firms are credit entities and investment firms which are members of the stock exchange and are entitled to trade in the market.
The firms trading in this market include banks, investment banks, financial management companies and brokerage firms among others. We empirically show that, although a firm may act on behalf of many individuals and institutions having different strategies, firms self-organize in groups with various degrees of specialization to the extent that in most cases it is possible to characterize a firm with a specific resulting strategy. We use a classification of firms in three groups to study the set of time intervals which are statistically indicating a collective action of the firms of a group with respect to the inventory variation of a given stock. In this way we are able to detect distinct herding characteristics of groups of firms characterized by different strategies. Our results are consistently observed over a four-year time period and can be used as empirical bases for an agent based models of financial markets.

{\it Inventory variation in a financial market} -- The activity of a market participant with respect to a given asset is well represented by the inventory variation which is the value exchanged as a buyer minus the value exchanged as a seller in a given time interval. In this article we investigate the inventory variation of financial firms exchanging a financial asset at the Spanish Stock Market (Bolsas y Mercados Espa\~noles, BME) during the years 2001 through 2004. In 2004 the BME was the eight in the world in market capitalization. Our database corresponds to the electronic open market SIBE (Sistema de Interconexi\'on Burs\'atil Electr\'onico) and allows us to follow each transaction performed by all the firms registered at BME. We have focused our 
investigation on Telef\'onica (TEF), Banco Bilbao Vizcaya Argentaria (BBVA), Banco Santander Central Hispano (SAN) and Repsol (REP) stocks, which are 4 highly capitalized stocks of BME and on the most active firms which have traded at least 200 trading days with at least 1000 transactions per year during the period 2001-2004. We investigate the market dynamics by focusing on the trading of each selected stock separately for each available calendar year. By doing so we have $4 \times 4$ distinct sets of results. The number of active firms is around $70$ with a minimum and maximum  value equal to $54$ and $82$, respectively. The homogeneity of obtained results for these sets is providing us indication about the general validity of them.

We first consider daily inventory variation of the investigated stock.  Let $v_i(t)$ indicate the inventory variation of firm $i$ during the day $t$.  For each investigated set, we have a multivariate time series of the inventory variation of the most active firms. We investigate the statistical properties of this set by considering its correlation coefficient matrix, which shows both positive and negative statistically significant correlation coefficients $\rho[v_i(t),v_j(t)]$. To assess if the detected correlations are carrying information about the market dynamics we perform a one factor or principal component analysis in which filtering of the components is done with the help of methods \cite{Laloux1999} based on random matrix theory (RMT). Particular attention has to be paid to spurious correlation due to the buy-sell counterparts present in each transaction. Even taking into account this consideration, we find that the first eigenvalue of the correlation matrix is not compatible with the null hypothesis of uncorrelated random inventory variables (see Fig.\ 1) and is therefore carrying information about the collective dynamics of firms. 

\begin{figure}
\includegraphics[scale=0.35,angle=-90]{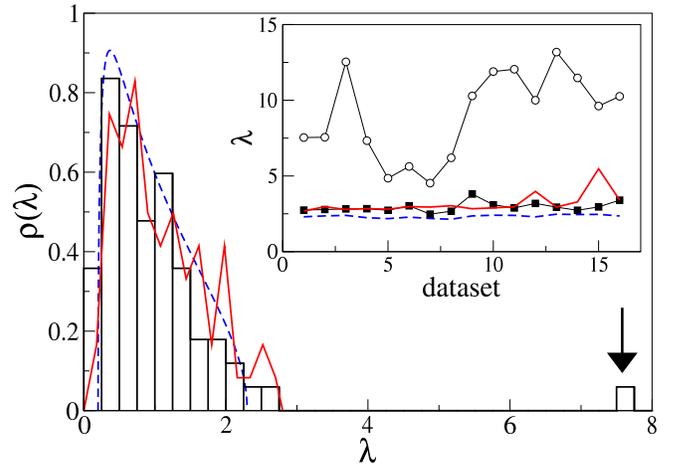}
\caption{Histogram (rectangles) of the eigenvalue spectrum of the correlation matrix of inventory variation of firms trading the stock BBVA in $2001$. The black arrow indicates the first eigenvalue. The blue dashed line is the spectral density expected by the Random Matrix Theory \cite{Laloux1999} where each time series is replaced by an uncorrelated random inventory time series. The solid red line is the averaged spectral density obtained by shuffling independently the buyers and the sellers, in such a way to maintain the same number of purchases and sales for each firm as in the real data. Other shuffling experiments give similar results.  In the inset we show the first (empty circles) and second (filled squares) eigenvalue of the 4 x 4 investigated sets. The dashed blue line again indicates the threshold expected by the RMT, and the solid red line is the upper threshold expected by the shuffling experiment. The first eigenvalue is well above the thresholds obtained with RMT and shuffling methods for all the investigated sets.}
\label{fig1}
\end{figure}

To elucidate the nature of this information we investigate the 
time profile of the factor associated with the first eigenvalue \cite{Mardia1979}: we find that there is a statistical significant correlation of this factor with the price return time series (see figure 2a and 2b) suggesting that the dynamics of the inventory variation can be succesfully described by the one factor model $v_i(t) = \gamma_i r(t) +\varepsilon_i(t)$ where $\gamma_i$ is proportional to the correlation between price return and inventory variation $ \rho[v_i(t),r(t)]$ and $\varepsilon_i(t)$ is a zero mean white noise term describing the idiosyncratic part of the strategy of the firm. We find that many firms are  significantly correlated (or anticorrelated) with price return and that correlations with price return are heterogeneous and depend on specific features of the trading firms like, for example, its size. Figure 2c shows that there are at least two groups of firms which are quite distinct with respect to their trading action. The firms characterized by a positive value of  $\gamma_i$  greater than the $2\sigma$ statistical uncertainty are categorized by us as trending firms. Conversely, when the correlation $\rho[v_i(t),r(t)]$ is negative and less then the $-2\sigma$ statistical uncertainty, the firms are categorized as reversing firms. We address the remaining firms as the uncategorized ones. Table I indicates that about 50\% of the firms are reversing whereas firms with trending strategy are observed approximately in the 10\% of the cases. The rest, around 40\%, remains uncategorized. Finally, Figure 2c also indicates a significant correlation between the strategy and size of the firm: in particular we find that BME is composed by few large trending firms and many reversing firms with a very heterogeneous size. As Table I indicates, the percent categorization over the four years is rather stable but what about to the behavior of a specific firm? In other words which is the
probability that a firm categorized in a given group will remain in the same
group or will move to some other group next year? We have computed the probability $P(Y|X)$ of a firm being in group $X$ in a given year and moving to group $Y$ during the next year. We have averaged these probabilities over the 3 changes of year present in our database. For the group of reversing firms ($X=R$), these probabilities are $P(R|R)=71\%$, $P(U|R)=16\%$, $P(T|R)=2\%$ and $P(E|R)=10\%$, where T indicates trending firms, U uncategorized ones and E indicates that the firm has exit from the set of active firms.
For the trending group we analogously obtain $P(R|T)=3\%$, $P(U|T)=34\%$, $P(T|T)=44\%$ and $P(E|T)=18\%$ whereas for the uncategorized firms we estimate $P(R|U)=19\%$, $P(U|U)=62\%$, $P(T|U)=7\%$ and $P(E|U)=12\%$. These probabilities show that a firm usually tends to preferentially stay in the same group over the years indicating a long term specialization. This behavior is more pronounced for reversing firms ($P(R|R)=71\%$) and less pronounced for trending firms ($P(T|T)=44\%$). Uncategorized firms are showing an intermediate behavior. The probability to move from reversing to trending firms, or viceversa, is rather low suggesting that such a change might occur only through successive changes which include a period of time spent in the uncategorized group. From this last group firms may move to trending and reversing group of firm with a probability $P(R|U)=19\%$ and $P(T|U)=7\%$ respectively. These results, obtained for Telef\'onica, are representative of the other stocks of our sample.
In summary a specialization of trading firms is present and rather stable over the years although some dynamics among groups is observed from year to year.

\begin{figure}[ptb]
\includegraphics[scale=0.5]{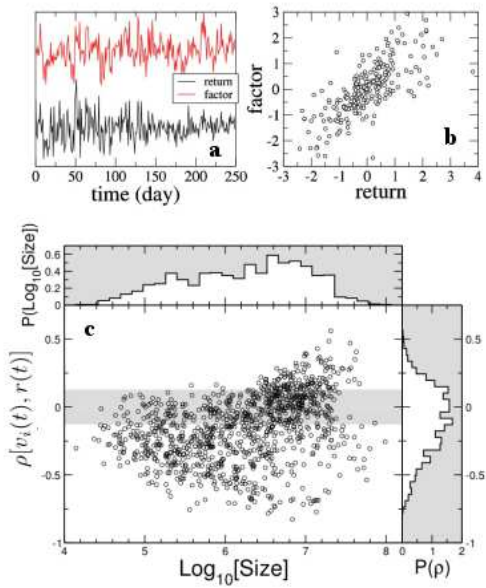}
\caption{Panel a) shows the time evolution of the first 
factor (red line) of the correlation coefficient matrix 
of daily inventory variation of firms trading the stock BBVA in 2003 
and the daily stock return of the same stock (black line).
In Fig. 2b we show the scatter plot of these quantities. A high degree 
of correlation (in the figure equals to 0.72) is observed. This result is
representative of all the investigated
sets where the factor - price return correlation is ranging from 0.47 to 0.74.
In Fig. 2c we show the scatter plot of the correlation between firm inventory variation and price return of the traded stock versus a proxy of the size of the firm. This proxy is the average of the total money exchanged daily by the firm. Size values are given in Euro. Each circle describes a firm trading a specific stock in a specific year.
The figure shows that both positive and negative correlation between inventory change and price return are present in the market. The gray rectangle points out the values
of $\rho[v_i(t),r(t)]$ which are within statistical uncertainty of $2\sigma$.
In the figure we also show the marginal probability density function 
of the correlation coefficient $\rho[v_i(t),r(t)]$ and of the size in the side panels of the figure.}
\label{fig2}
\end{figure}

The price return $r(t)$ of the traded stock acts as a common factor for all the firms. Our one-factor model predicts that the cross correlation $\rho[v_i(t),v_j(t)]$ between the inventory variation of two firms is significantly positive when both firms $i$ and $j$ are trending or reversing firms, whereas the cross correlation is significantly negative when the two firms belong one to the first and the the other to the second group. When one of the firms belongs to the third group the cross correlation is not significantly different from zero. 
In order to illustrate how well the model reproduce the empirical data, we 
show in Fig 3 the contour plot of the correlation matrix of daily inventory 
variation plotted by sorting the firms in the rows and columns according to 
their value of correlation $\rho[v_i(t),r(t)]$. The approximately block structure of the matrix indicates that the proposed model gives a good basic description of data.
Moreover it can be shown \cite{Lillo2005}, that the correlation 
matrix of the model is composed by a large eigenvalue 
$\lambda_1\sim\sum\gamma_i^2$ and $N-1$ small eigenvalues, similarly to what 
is seen in Fig. 1 for empirical data.

\begin{table}
\caption{Number of active firms belonging to the groups of reversing, uncategorized and trending firms for the calendar years of the period 2001-2004.
The firms are active firms trading the Telef\'onica stock. In parenthesis we report the percentage of firms in the considered group for each year.}
%\begin{center}
\begin{tabular}{l||c|c|c|c|}
~~~ & 2001 & 2002 & 2003 & 2004 \\
\tableline
Reversing & $ 43 ~~~ (52\%) $ & $ 39 ~~~ (49\%) $ & $ 42 ~~~ (52\%) $ & $ 37 ~~~ (51\%) $ \\ 
Uncategorized & $ 28 ~~~ (34\%) $ & $ 31 ~~~ (39\%) $ & $ 31 ~~~ (38\%) $ & $ 29 ~~~ (40\%) $ \\ 
Trending & $ 11 ~~~ (13\%) $ & $ 10 ~~~ (12\%) $ & $ 8 ~~~ (9.9\%) $ & $ 6 ~~~ (8.3\%) $ \\ 
Total & $ 82 ~~~  $ & $ 80 $ & $ 81 $ & $ 72 $ \\ 
\tableline
\end{tabular}
% \end{center}
\label{summary}
\end{table}

\begin{figure}
\includegraphics[scale=0.45]{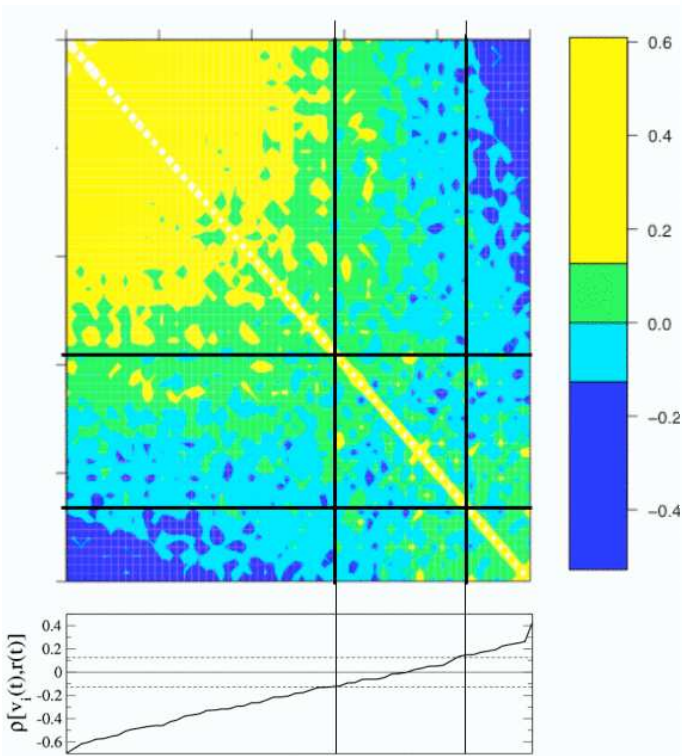}
\caption{Contour plot of the correlation matrix of daily inventory 
variation of firms trading the stock BBVA in 2003 plotted by 
sorting the firms in the rows and columns according to 
their value of correlation of inventory variation with BBVA price return $\rho[v_i(t),r(t)]$. 
The bottom panel shows the value of $\rho[v_i(t),r(t)]$ of the firms in the same order as in the matrix. The dashed lines in the bottom panel bounds the $2\sigma$ significance interval. 
Colors of the matrix are chosen to highlight positive and negative firm daily inventory 
cross correlation values $\rho[v_i(t),v_j(t)]$ above a given significance level. Specifically, yellow (blue) indicates positive (negative) cross correlation with a significance of $2\sigma$, whereas green (cyan) indicates positive (negative) cross correlation below $2\sigma$.   
Two groups of firms are seen, one on the top left corner and the other on the bottom right corner.
It should be noted that these two groups of firms present a significant level of anticorrelation of their inventory variation profile (the blue areas observed at the left-top and right-bottom corners of the matrix). The thick black lines in the matrix are obtained from the bottom panel by partitioning the firms in three groups according to the value of $\rho[v_i(t),r(t)]$ (smaller than $-2\sigma$, between $-2\sigma$ and $2\sigma$ and larger than $2\sigma$). }
\label{fig3}
\end{figure}

The correlation between inventory variation and price return raises the question of the causality relation between these two variables.
This problem has been investigated in ref. \cite{Griffin2003} by studying a specialized proprietary database of institutions and individuals trading in Nasdaq 100 stocks from 1/2000 to 2/2001. Here we investigate the same problem for the two groups of firms characterized by a trending or reversing resulting strategy. Specifically, we investigate inventory variation
and price return time series sampled
at a 15 minute time horizon. The high frequency autocorrelation 
of price return shows that linear autocorrelation has a time scale
strictly shorter that 15 minutes therefore confirming a weak form of
market efficiency \cite{Campbell1997}. Trending firms are characterized by an autocorrelation of $v(t)$ which is significantly positively correlated up to a few trading hours.
Conversely, reversing firms
have $v_i(t)$ negatively autocorrelated on a time scale of 15 minutes.
Thus few large trending firms act on a long time scale in order to build a position by splitting large orders \cite{Chan1995,Gabaix2003,Lillo2004,Vaglica2007} 
to minimize their market impact \cite{Hasbrouck1991}, whereas a population of many reversing firms of different size primarily act on a short time scale. 

The investigation of the high frequency lagged cross correlation between $r(t)$ and $v_i(t)$ provide us information about the causality direction between the two variables. 
Fig. 4 shows that inventory variation in the future $v_i(t)$ is correlated with price return in the near past $r(t+\tau)$ ($\tau < 0$), whereas $v_i(t)$ is not significantly correlated with $r(t+\tau)$ in the near future ($\tau > 0$). Moreover,  as shown in figure 4, the larger is the synchronous correlation $\rho[v_i(t), r(t)]$, the larger is the cumulated effect on future inventory variation $\sum_{\tau<0} \rho[v_i(t),r(t+\tau)] $. However, since there is no significant correlation between the present inventory variation and the future return, the cumulated effect disappears for $\tau > 0$. This result indicates that firms of both groups are ``following" the price (in different directions) rather than pushing it in a given direction. 

\begin{figure}
\includegraphics[scale=0.5]{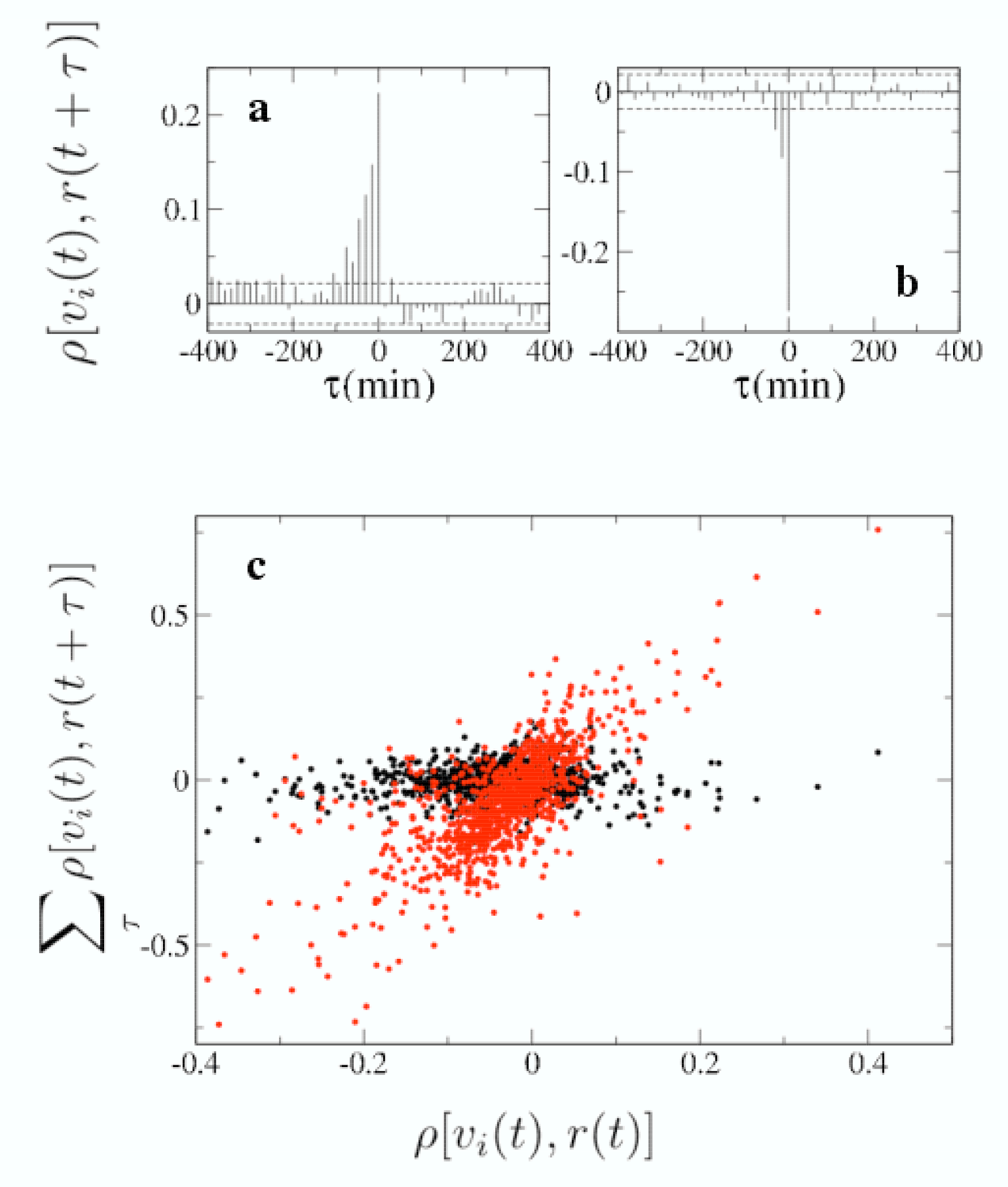}
\caption{Panel a) and b) show the lagged cross correlation $\rho[v_i(t),r(t+\tau)]$ for two firms representative of the trending and reversing groups respectively. The dashed lines bounds $2\sigma$ significance interval. There is a significant lagged cross correlation up to 75 minute for the trending firm and to 30 minute for the reversing firm. The lagged cross correlation values show that $r(t)$ is correlated with inventory variation in the near future $v_i(t+\tau)$, but not the other way around.
In panel c), each circle represents the integrated lagged cross correlation between inventory variation and return. For red circles, we integrate the lagged cross correlation as $\sum \rho[v_i(t), r(t+\tau),] $ between $\tau=-15$ to $\tau=-150$ minutes and we plot the integrated cross correlation versus the simultaneous cross correlation $\rho[v_i(t),r(t)]$. For black circles, we integrate the lagged cross correlation as $\sum \rho[v_i(t), r(t+\tau),]$ between $\tau=15$ to $\tau=150$ minutes. Causality is detected for the first set of points and absent in the second one.}
\label{fig4} 
\end{figure}

{\it Degree of common behavior of group of firms}.

Are firms belonging to the same group behaving in a similar way at specific time intervals? To answer this question we use an indicator based on the inventory variation of each firm. The herding indicator 
\begin{equation}
h=\frac{\# ~ of ~ buying ~ firms}{\# ~ of ~ buying ~ firms +\# ~ of ~ selling ~ firms} 
\end{equation}
of the group is the number of buying firms divided by the number of firms of the group which are active in the specific time interval (buying or selling). This herding indicator is a simplified version of the herding measure introduced in Ref. \cite{Lakonishok1992} to quantify the herding of institutional investors in selecting a basket of stocks. Differently than in Ref. \cite{Lakonishok1992} here we limit our investigation to the univariate case of the investment in a single stock. We infer that herding is associated to the observation of a high value (buy herding) or low value (sell herding) of $h$ by evaluating the probability to observe a number of buying (selling) firms equal or larger than the empirically detected one under a binomial null hypothesis. Specifically, we infer that herding is present when the probability of the observed number of buying or selling firms is smaller than $5\%$ under the null binomial hypothesis.

We estimate $h$ for the three groups of stocks at different time horizons ranging from 15 minutes to 5 trading days. A selection of the results obtained are summarized in Table~\ref{IntTef} for the one day and 15 minutes time horizon respectively. By analyzing Table~\ref{IntTef} we note that firms characterized by a reversing resulting strategy present herding in a significant fraction of time intervals. Specifically the percent of herding intervals averaged over four years ranges from 32.6\% for the 15 minutes time horizon to 66.4\% for the 1 day time horizon. The percent of herding time intervals is much less pronounced for firms with a trending resulting strategy. For this group we observe a percent of herding intervals of a few percent for the time horizon both of 15 minutes (4.3\%) and 1 trading day (6.3\%). The uncategorized firms present an intermediate behavior.

\begin{widetext}
\begin{center}
\begin{table}[t]
\caption{Percent of herding intervals observed for the groups of reversing, uncategorized and trending firms actively trading the Telef\'onica stock during the period 2001-2004. The percent of herding intervals is also provided separately for buying (BH) and selling (SH) herding. The first three lines refer to the 1 day time horizon. For each calendar year the total number of trading days is 250. The last three lines refer to the 15 minutes intraday time horizon. The total number of 15 minutes intervals is 8500 for each calendar year.}
%\begin{center}
\begin{tabular}{l||c|c|c|c|}
~~~ & 2001 & 2002 & 2003 & 2004 \\
~~~ & BH~~|~~SH & BH~~|~~SH & BH~~|~~SH & BH~~|~~SH \\
\hline
Reversing &  68.0 \% & 72.4 \% & 64.0 \% & 61.2 \% \\ 
(1 d) &  40.8\%~~|~~27.2\% & 50.0\%~~|~~22.4\% & 29.2\%~~|~~34.8\% &  23.6\%~~|~~37.6\% \\
\hline
Uncategorized &  23.2 \% & 18.4 \% & 24.8 \% & 20.8 \% \\ 
(1 d) &  16.4\%~~|~~6.8\% & 13.6\%~~|~~4.8\% & 10.8\%~~|~~14.0\% &  6.8\%~~|~~14.0\% \\
\hline
Trending &  10.4 \% & 6.4 \% & 6.0 \% & 2.4 \% \\ 
(1 d) &  7.2\%~~|~~3.2\% & 2.4\%~~|~~4.0\% & 2.0\%~~|~~4.0\% &  1.2\%~~|~~1.2\% \\
\hline
\hline
Reversing &  36.2 \% & 37.3 \% & 29.2 \% & 27.7 \% \\
(15 m) &  21.7\%~~|~~14.5\% & 24.9\%~~|~~12.4\% & 13.4\%~~|~~15.8\% &  10.8\%~~|~~16.9\% \\
\hline
Uncategorized &  10.7 \% & 12.4 \% & 10.8 \% & 13.7 \% \\
(15 m) &  7.3\%~~|~~3.4\% & 7.6\%~~|~~4.8\% & 3.8\%~~|~~7.0\% &  3.5\%~~|~~10.2\% \\
\hline
Trending &  3.3 \% & 6.6 \% & 3.9 \% & 3.3 \% \\
(15 m) &  2.1\%~~|~~1.2\% & 3.4\%~~|~~3.2\% & 1.7\%~~|~~2.2\% &  1.7\%~~|~~1.6\% \\
\hline
\hline
\end{tabular}
% \end{center}
\label{IntTef}
\end{table}
\end{center}
\end{widetext}

\begin{figure}
\includegraphics[scale=0.35]{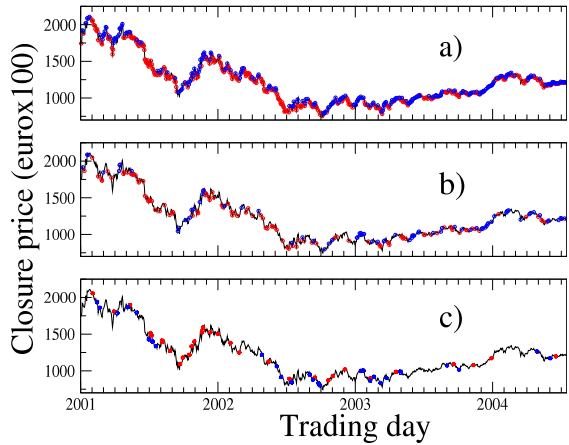}
\caption{The thin black line is the daily closure price of Telef\'onica stock for the January 2001 - December 2004 time period. The three panels refer to reversing (a), uncategorized (b) and trending (c) firms. Red circles indicate buying herding days whereas blue circles indicate selling herding days. The percent of herding days is 66.4\%, 21.8\% and 6.3\% for reversing, uncategorized and trending firms respectively.}
\label{herding} 
\end{figure}

An illustration of the occurrence of the herding time intervals estimated for the one day time horizon is provided in Fig. \ref{herding} where we show the time evolution of daily closure stock price of Telef\'onica during the time period 2001-2004. Each panel refers to a different group of firms. The herding days of reversing firms are highly frequent and approximately uniformly distributed over the investigated time period. The prevalence over long periods of time of the kind of observed herding (buying or selling) is related to the prevalence of a bull or bear market phase.  The herding days detected for the trending firms are much less frequent and their occurrence is less uniformely distributed in time.

The herding measure we use is deliberately simple and it needs additional validation to prevent the possibility that the detected herding intervals  do not  imply a significant change of the group inventory variation. In fact, for example, we could imagine the possibility that several firms are just net buying each a small amount of value of the traded stock whereas the remaining few firms are selling the traded stock for an amount corresponding to a much larger value. In this case, with our methodology we would infer buy herding in the presence of a net sell from the considered group of firms. We therefore need to complement our herding estimation with additional indicators. We quantify the degree of common action of firms in a group with respect to the exchanged value of the traded stock by considering the buy ratio $b$ used by Grinblatt and Keloharju  in their study of investment behavior of private and institutional Finnish investors \cite{Grinblatt2000}. Specifically, for each time interval and for each group we compute
\begin{equation}
b=\frac{\sum_{i~\in~ buying firms} v_i}{\sum_{i~\in~buying firms} v_i+\sum_{i~\in~selling firms} |v_i|}
\end{equation}
The buy ratio $b$ is varying between zero and one. Low values of $b$ are indicating time intervals when firms of the considered group are mostly selling whereas high values close to one are indicating that firms are mostly buying. In addition to the buy ratio we also consider a last indicator used to quantify the degree of activity of various firms belonging to each group in a specific time interval. This is done by adapting the ``effective" number indicator originally proposed to quantify the amount of stocks significantly present in a portfolio \cite{BPbook}. We consider
\begin{equation}
N_{eff}=\frac{1}{\sum_{i=1}^N \omega_i^2}
\end{equation}
where $\omega_i=v_i/\sum_{i=1}^N |v_i|$ and $N$ is the number of firms in the considered group. This indicator is ranging from the value $N_{eff}=1$ when only one firm is responsible for the group inventory variation to the value $N_{eff}=N$ which is indicating that all firms are contributing equally to the inventory variation of the group of firms.

\begin{widetext}
\begin{center}
\begin{table}
\caption{Mean value of the buy ratio $b$ and of the fraction of the number of effective firms $N_{eff}/N$ which are active in a given time interval. The mean values are computed both unconditional and conditional on the buying (BH) and selling (SH) herding intervals for the reversing, uncategorized and trending groups of active firms trading the Telef\'onica stock in 2001. The first three lines refer to the 1 day time horizon whereas the last three lines refer to the 15 minutes time horizon. The dispersion of mean value is plus or minus one standard deviation. Number in parenthesis is the number of records of the considered set. The presence of symbols $\oplus$, $\triangle$, $\Box$ and $\Diamond$ for each group of firms and for each time horizon select the pairs of unconditional and conditional mean values which does not pass a $t$-test assuming the null hypothesis that the two sets of data providing unconditional and conditional mean values originates from the same normal statistics. These pairs of mean values are statistically distinct at a 99\% confidence level.}
%\begin{center}
\begin{tabular}{l||c|c|c|c|c|c}
~~~ & $<b>$ & $<b>_{BH}$ & $<b>_{SH}$ & $<N_{eff}/N>$ & $<N_{eff}/N>_{BH}$ & $<N_{eff}/N>_{SH}$\\
\hline
Reversing &  $0.52\pm0.28$  & $0.75\pm0.16$ & $0.20\pm0.14$ & $0.19\pm0.06$ & $0.20\pm0.07$ & $0.20\pm0.06$ \\ 
(1 d) & $\oplus$ $\triangle$ (250) & $\oplus$ (102) & $\triangle$ (68) & (250) & (102) & (68) \\
\hline
Uuncategorized &  $0.48\pm0.16$  & $0.55\pm0.16$ & $0.40\pm0.14$ & $0.26\pm0.07$ & $0.26\pm0.07$ & $0.25\pm0.07$ \\ 
(1 d) &  (250) & (41) & (17) & (250) & (41) & (17) \\
\hline
Trending &  $0.51\pm0.25$  & $0.81\pm0.20$ & $0.22\pm0.19$ & $0.41\pm0.12$ & $0.44\pm0.12$ & $0.37\pm0.08$ \\ 
(1 d) &  $\oplus$ $\triangle$ (250) & $\oplus$ (18) & $\triangle$ (8) & (250) & (18) & (8) \\
\hline
\hline
Reversing &  $0.50\pm0.22$  & $0.67\pm0.18$ & $0.31\pm0.18$ & $0.15\pm0.06$ & $0.16\pm0.06$ & $0.16\pm0.06$ \\ 
(15 m) & $\oplus$ $\triangle$ (8500) & $\oplus$ (1845) & $\triangle$ (1230) & $\Box$ $\Diamond$ (8500) & $\Box$ (1845) & $\Diamond$ (1230) \\
\hline
Uncategorized &  $0.49\pm0.23$ & $0.68\pm0.21$ & $0.29\pm0.20$ & $0.16\pm0.06$ & $0.18\pm0.06$ & $0.18\pm0.06$ \\ 
(15 m) & $\oplus$ $\triangle$ (8500) & $\oplus$ (622) & $\triangle$ (293) & $\Box$ $\Diamond$ (8500) & $\Box$ (622) & $\Diamond$ (293) \\
\hline
Trending &  $0.51\pm0.25$  & $0.88\pm0.19$ & $0.15\pm0.20$ & $0.31\pm0.10$ & $0.31\pm0.12$ & $0.32\pm0.11$ \\ 
(15 m) & $\oplus$ $\triangle$ (8500) & $\oplus$ (181) & $\triangle$ (102) &  (8500) & (181) & (102) \\
\hline
\hline
\end{tabular}
%\end{center}
\label{Validation}
\end{table}
\end{center}
\end{widetext}

To validate the herding behavior observed in Table~\ref{IntTef} we present in Table~\ref{Validation} the mean value of $b$ and $N_{eff}/N$ for all the groups and for the same values of time horizon used in Table~\ref{IntTef}. The results refers to the active firms trading the Telef\'onica stock during 2001. A similar behavior is observed for the other investigated years. The values of $b$ and $N_{eff}/N$ are computed both unconditional on all the investigated time intervals and conditioning on the intervals characterized as buying or selling herding intervals by the herding indicator with the associated binomial test.

Table~\ref{Validation} shows that the mean value of $b$ is, as expected, very close to 0.5 when the average is performed unconditionally over all intervals. Differently, when the average is computed conditioning on buying or selling herding intervals (as detected by $h$ of Eq. (1)) one obtains mean values of $b$ higher or smaller than 0.5 respectively. The Table also shows the values of 
$<N_{eff}/N>$ computed both unconditional and conditioning over the herding intervals. 
To assess the statistical reliability of the differences observed between the unconditional mean values and the conditional ones we perform a $t$-test at a 99\% confidence threshold of the null hypothesis that the unconditional and conditional means both originates from the same set of normal data. In Table~\ref{Validation} we label with symbols $\oplus$ and $\triangle$ the rejection of this null hypothesis when we compare the unconditional mean of the buy ratio $b$ with the one obtained conditioning to buying ($\oplus$) and selling ($\triangle$) herding days. Analogously, we label with symbols $\Box$ and $\Diamond$ the rejection of the null hypothesis related to the the comparison between the unconditional mean of $N_{eff}/N$ and the mean estimated conditioning to buying ($\Box$) and selling ($\Diamond$) herding days. 

The results summarized in Table~\ref{Validation} show that the differences between the unconditional mean values of the buy ratio $b$ and the conditional ones are statistically significant in all cases with the important exception of the uncategorized group investigated at the one day time horizon. Therefore for reversing and trending firms the buy ratio observed during herding is different from 0.5 in a statistical significant way and it is fully supporting the herding hypothesis.
For the values of $N_{eff}/N$ we observe that a statistically robust difference between the unconditional mean values and the conditional ones is observed only for a 15 minutes time horizon for the reversing and uncategorized group. However, also in these cases the difference between the unconditional and conditional values is almost negligible. In fact for this variable, in all cases, the values computed conditionally are very close to or statistically indistinguishable from the values computed unconditionally. This result indicates that during herding intervals the relative activity $\omega_i$ of various firms of each group is, on average, not too far from the unconditional value reflecting the size heterogeneity of the group.  By summarizing the results obtained about $<b>$ and  $<N_{eff}/N>$ and their statistical validation we conclude that our herding indicator selects herding intervals which are in average characterized by values of a buy ratio which are supporting the herding hypothesis. We also note that the set of herding intervals is characterized by the same relative activity of the firms of the three groups observed unconditionally.

The values of $<N_{eff}/N>$ of Table~\ref{Validation} show that $<N_{eff}/N>$ is higher for a longer time interval. Moreover, for each time interval $<N_{eff}/N>$ is higher for trending  firms. This result is consistent with the observation that the group of firms with a resulting trending strategy is the most homogeneous one with respect to the size of firms as shown in Fig.~\ref{fig2}c.

Our herding indicator is therefore validated and the results obtained show that the herding behavior of firms with a resulting reversing strategy is markedly different from the herding behavior of firms with a trending strategy.

{\it Conclusion} -- Our results show that a large number of firms trading a financial asset in a financial market are characterized by a well defined resulting strategy. Specifically we detect financial firms that can be classified as trending or reversing firms. In the Spanish Stock Market trending firms are few large financial firms whereas many firms, which are heterogeneous in size, are reversing firms. These findings indicate that most financial firms highly specialize their trading activity. Each resulting strategy most probably reflects the most common or the most prominent investment strategy of major clients of the financial firm. 

We also show that trending and reversing firms present a caracteristic pattern of herding behavior both at daily and at intradaily time horizons. Reversing firms are herding quite frequently and uniformly in time whereas trending firms are herding rarely. However, when herding is present the involvement of firms of the group is more pronounced and uniform in trending rather than in reversing firms.  

Market dynamics can therefore be seen as the interplay of at least two classes of traders, different in size and responding to the price changes in different ways. It is possible that the fluctuation of price returns, i.e. the market volatility, is significantly affected by the fluctuations in the relative trading intensity of the two groups. Our results open up the possibility of setting up agent based models of financial firms trading in a financial market. These models can now be empirically grounded in the type of resulting strategies characterizing the dynamics of real firms.

\medskip

{\bf Acknowledgments} Authors acknowledge Sociedad de Bolsas for providing the data and the Integrated Action Italy-Spain ``Mesoscopics of a stock market" for financial support. 
F.L., G.V., and R.N.M. acknowledge partial support from MIUR research project ``Dinamica di altissima frequenza nei mercati finanziari",and NEST-DYSONET 12911 EU project. E.M. acknowledges partial support from MEC (Spain) throught grant FIS2004-01001, MOSAICO and a Ram\'on y Cajal contract and Comunidad de Madrid through grants UC3M-FI-05-077 and SIMUMAT-CM.

\end{document}